\documentclass[prb,twocolumn,superscriptaddress,showpacs,amsmath,amssymb]{revtex4-2}
\usepackage[colorlinks=true, linkcolor=blue, filecolor=blue, urlcolor=blue, citecolor=blue]{hyperref}
\usepackage{amsfonts}
\usepackage{graphicx}

\begin{document}

\title{Helical Fermi Arc in Altermagnetic Weyl Semimetal}

\author{Yu-Hao Wan}
\affiliation{International Center for Quantum Materials , School of Physics, Peking
University, Beijing 100871, China}

\author{Cheng-Ming Miao}
\affiliation{International Center for Quantum Materials , School of Physics, Peking
University, Beijing 100871, China}

\author{Peng-Yi Liu}
\affiliation{International Center for Quantum Materials , School of Physics, Peking
University, Beijing 100871, China}

\author{Qing-Feng Sun}
\email[Corresponding author: ]{sunqf@pku.edu.cn}
\affiliation{International Center for Quantum Materials , School of Physics, Peking
University, Beijing 100871, China}
\affiliation{Hefei National Laboratory, Hefei 230088, China}

\begin{abstract}
We investigate the topological properties of modified Dirac Hamiltonians
with an altermagnetic mass term and reveal a novel mechanism for realizing
altermagnetic Weyl semimetals. Unlike the conventional Wilson mass,
the altermagnetic mass drives direct transitions between nontrivial
Chern phases of opposite sign and fundamentally reshapes the band
inversion surface. By extending this framework to three dimensions,
we construct a minimal lattice model that hosts pairs of Weyl nodes
as well as coexisting helical Fermi arcs with opposite chirality on
the same surface, which is a phenomenon not found in conventional
magnetic Weyl semimetals. We further propose a practical scheme to realize
these phases in multilayer structures of 2-dimensional Rashba metal with
engineered $d$-wave altermagnetic order. Our results deepen the theoretical
understanding of mass terms in Dirac systems and provide concrete
guidelines for the experimental detection and realization of altermagnetic
Weyl semimetals.
\end{abstract}
\maketitle

\section{Introduction}\label{I}

Weyl semimetals (WSMs) are gapless three-dimensional topological
phases characterized by isolated band-touching points known as Weyl
nodes, which act as quantized monopoles of Berry curvature \cite{wan_topological_2011,armitage_weyl_2018}.
Realizing WSM phases typically involves breaking either inversion ($\mathcal{P}$) \cite{shuichi_murakami_phase_2007,halasz_time-reversal_2012,PhysRevB.108.235211}
or time-reversal $(\mathcal{T})$ \cite{burkov_weyl_2011,liu_giant_2018} symmetry, positioning magnetism as a natural and experimentally accessible route to achieve and control these states.
To date, most of the magnetic WSMs rely on ferromagnetic \cite{wang_time-reversal-breaking_2016,liu_giant_2018,nie_magnetic_2022,PhysRevB.110.125204} or antiferromagnetic \cite{kuroda_evidence_2017,grassano_type-i_2024} order, whose net or staggered magnetization explicitly breaks $\mathcal{T}$, producing characteristic responses such as anomalous Hall effects \cite{burkov_anomalous_2014,burkov_chiral_2015,itoh_weyl_2016,kuroda_evidence_2017,liu_giant_2018}, chiral anomalies \cite{nielsen_adler-bell-jackiw_1983,aji_adler-bell-jackiw_2012,liu_chiral_2013,burkov_chiral_2015,aref1,aref2}, and surface Fermi-arc states \cite{wan_topological_2011,morali_fermi-arc_2019,ma_linked_2021,zheng_observation_2022}.

A newly recognised form of collinear order, altermagnetism, now enlarges
this landscape \cite{vsmejkal_crystal_2020,krempasky_altermagnetic_2024,song_altermagnets_2025,cjzw-j4v7}.
An altermagnet has zero net magnetization yet still
breaks $\mathcal{T}$ symmetry in momentum space, generating significant
spin splittings strongly dependent on crystal symmetry \cite{vsmejkal_beyond_2022,vsmejkal_emerging_2022}.
The unique properties of altermagnets have attracted substantial theoretical and experimental attention due to their potential for unconventional spintronic effects \cite{bai_observation_2022,karube_observation_2022,vsmejkal_giant_2022}, anomalous transport phenomena \cite{vsmejkal_crystal_2020,gonzalez-hernandez_efficient_2021,yi_spin_2025,wan_altermagnetism-induced_2025,chen2025anomalous,aref3,new1,new2,PhysRevB.111.165421,3hxt-fynf}, and novel superconducting effects \cite{vsmejkal_emerging_2022,ouassou_dc_2023,zhu_topological_2023,cheng_orientation-dependent_2024,cheng_field-free_2024,banerjee_altermagnetic_2024,sc1}.
However, the impact of altermagnetism on Weyl physics remains
largely unexplored. In particular, can altermagnetic order stabilize
Weyl phases distinct from those realized in ferromagnetic or antiferromagnetic
systems?

To address this, it is essential to understand how different mass
terms in the modified Dirac equation shape the topological phases
of both two- and three-dimensional systems \cite{shen_topological_2012}.
While most studies have focused on the interplay between the uniform Dirac mass and the quadratic Wilson mass \cite{qi_topological_2006,bernevig_quantum_2006,zhang_topological_2009}, recent work suggests that symmetry-allowed $d$-wave altermagnetic mass terms may fundamentally alter the topological phase diagram\cite{wannew1}, enabling novel WSM phases with properties absent in ferromagnetic or antiferromagnetic systems.

In this work, we show that combining Dirac and altermagnetic masses
indeed yields a distinct class of Weyl semimetal. Using a minimal
square-lattice model we demonstrate that Weyl nodes arise precisely
where the two masses cancel, and that their arrangement differs qualitatively
from the Wilson-mass case. Most strikingly, the resulting phase hosts
two counter-propagating Fermi arcs on the same surface, forming a
helical surface channel absent in conventional magnetic WSMs. We then
propose a multilayer structure, where two-dimensional metallic layers
with $d$-wave altermagnetic order and alternating Rashba spin-orbit
coupling are stacked with insulating spacers. This setup naturally
realizes the required mass structure and offers a platform
for studying the predicted effects. Our findings reveal a new route
to engineer and control unconventional topological phases through
altermagnetism, opening up opportunities for observing exotic surface
states and transport phenomena beyond those of traditional magnetic
Weyl semimetals.

The rest of the paper is organized as follows. In Sec. \ref{II}, we analyze
the differences between Wilson and altermagnetic masses in the two-dimensional
modified Dirac equation, highlighting their distinct topological implications.
In Sec. \ref{III}, we extend the model to three dimensions and demonstrate
how the interplay between the Dirac and altermagnetic masses gives
rise to altermagnetic Weyl semimetals with helical fermi arc surface
states. In Sec. \ref{IV}, we propose and analyze the multilayer heterostructure
and map out the corresponding topological phase diagram. Finally,
we summarize our findings in Sec. \ref{V}.

\begin{figure}
\begin{centering}
\includegraphics[width=\columnwidth]{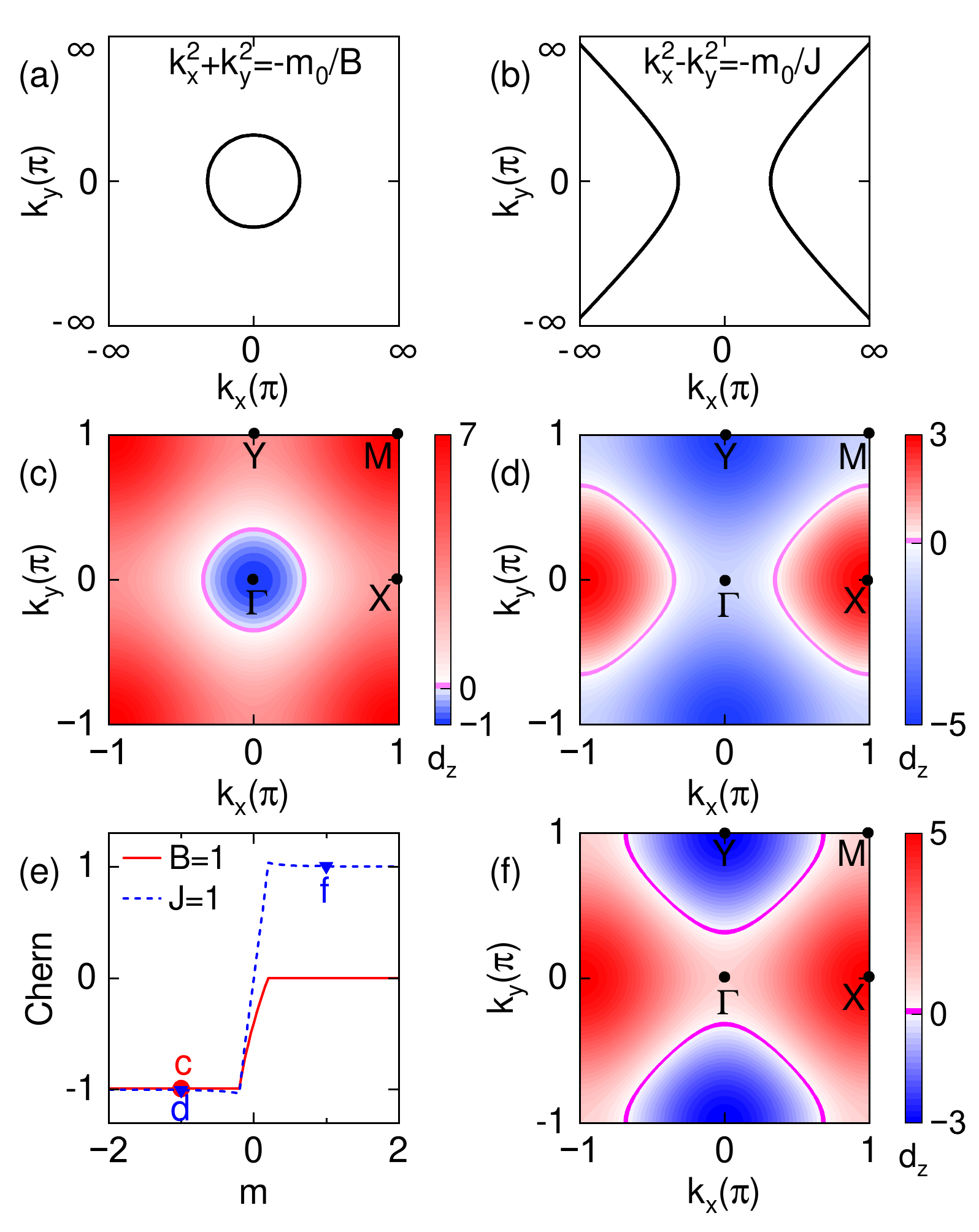}
\par\end{centering}
\caption{\label{fig:1}(a) BIS for the Dirac mass and Wilson mass in the continuum
model. (b) BIS for the Dirac mass and altermagnetic mass in the continuum
model. (c) BIS for the Dirac mass and Wilson mass in the lattice model, with $m_0=-1$ and $J=1$;
color indicates the $d_{z}$ component. (d) BIS for the Dirac mass
and altermagnetic mass in the lattice model, with $m_0=-1$ and $J=1$. (e) Chern number as a function of Dirac mass $m_0$ for altermagnetic mass
(blue line) and Wilson mass (red line); the three points marked c, d, and f correspond
to different $d_{z}$ distributions in the Brillouin zone in the panels (c, d, and f). (f) BIS
for the Dirac mass and altermagnetic mass in the lattice model, with
$m_0=1$ and $J=1$.}
\end{figure}

\section{Topological Effects of Dirac, Wilson, and Altermagnetic Mass Terms}\label{II}

The rich topological structure of WSMs can be traced back to their deep connection with Chern insulators.
In particular, the modified Dirac Hamiltonian provides a universal minimal description of band topology and topological phase transitions in both two- and three-dimensional systems.
The key physics is encoded in the mass term $M(k)$, whose precise form determines the nature of the band inversion and the topological classification.

Most studies have focused on the interplay between a constant Dirac
mass and a quadratic Wilson mass, but recent advances in altermagnetic
systems suggest that an alternative quadratic mass term\textemdash associated
with $d$-wave altermagnetic order\textemdash may fundamentally reshape
the topological phase diagram. To clarify these distinctions and motivate
our construction of altermagnetic Weyl semimetals, we systematically
compare the topological effects of Dirac, Wilson, and altermagnetic
mass terms using the 2D modified Dirac model. In two dimensions, the
minimal model is given by
\begin{equation}
H(\mathbf{k})=v_{F}(k_{x}\sigma_{x}+k_{y}\sigma_{y})+M(\mathbf{k})\sigma_{z},
\end{equation}
where $v_{F}$ is the Fermi velocity, $\sigma_{x,y,z}$ are the Pauli
matrices in the spin space, and $\mathbf{k}=(k_{x},k_{y})$ is the crystal momentum.
The mass term $M(\mathbf{k})$ is central to determining the topological
nature of the system.

For the modified Dirac equation, the mass term typically contains
a constant Dirac mass $m_{0}$ and a quadratic Wilson mass $B(k_{x}^{2}+k_{y}^{2})$, where $B$ is the Wilson mass parameter \cite{kogut_hamiltonian_1975,rothe_lattice_2012,fu_quantum_2022}.
The interplay between the Dirac mass $m_{0}$ and the Wilson mass $B$ governs the system's topological phase when the product $m_{0}B<0$, the model describes a nontrivial Chern insulator, while $m_{0}B>0$ results in a trivial insulator \cite{qi_topological_2006,shen_topological_2012}.
The transition point corresponds to a band inversion, where
the energy gap closes and reopens as the sign of $m_{0}$ or $B$
changes.

Recently, motivated by the physics of $d$-wave altermagnetic order,
another type of quadratic mass term has been considered \cite{vsmejkal_beyond_2022,vsmejkal_emerging_2022,yi_spin_2025,wannew1}
\begin{equation}
M_{\mathrm{S}}(k_{x},k_{y})=J(k_{x}^{2}-k_{y}^{2}),
\end{equation}
where $J$ is the strength of the altermagnetic ordering.
Like the Wilson mass, this term is quadratic in momentum and couples to $\sigma_{z}$, but it breaks rotational symmetry, introducing a sign change between the $k_{x}$ and $k_{y}$ directions.

To illustrate the fundamental difference between these two mass terms,
we examine the geometry of the band inversion surface (BIS), defined
as the set of momentum points where the mass term vanishes and the
bulk gap closes. For the Dirac equation with a Wilson mass, the BIS
is determined by $k_{x}^{2}+k_{y}^{2}=-m_{0}/B$, forming a circle
centered at the origin [see Fig. \ref{fig:1}(a)]. In contrast, for
the Dirac equation with an altermagnetic mass, the BIS is given by
$k_{x}^{2}-k_{y}^{2}=-m_{0}/J$, which represents a hyperbola [see Fig.
\ref{fig:1}(b)]. These distinct BIS geometries reflect the underlying
symmetry of each mass term and dictate different mechanisms for gap
closing and topological phase transitions.

These geometric differences in the BIS have profound topological implications.
The circular BIS associated with the Wilson mass supports a well-defined
Chern number in the continuum, as the mass term maintains a uniform
sign at infinity. In contrast, the altermagnetic mass leads to a non-uniform
mass sign at large momenta, so the mapping from momentum space to
the Bloch sphere fails to close, rendering the Chern number ill-defined
in the pure continuum limit.

To rigorously define topological invariants, we regularize the model
on a lattice, where the Brillouin zone is compact. On the square lattice,
the Wilson mass term takes the form $m_{0}-2B(\cos k_{x}+\cos k_{y}-2)$.
Setting $B=1$, we compute the Chern number as a function of $m_{0}$.
As shown in Fig. \ref{fig:1}(e), the Chern number changes from -1 ($m_{0}<0$)
to 0 ($m_{0}>0$), reflecting a topological phase transition driven
by the sign reversal of the Dirac mass. Correspondingly, the BIS for
the Wilson mass is a closed loop centered at the $\Gamma$ point in
the Brillouin zone [see Fig. \ref{fig:1}(c)].

For the altermagnetic mass, the lattice regularization yields $2J(\cos k_{y}-\cos k_{x})$,
and we set $J=1$ for comparison. In this case, as $m_{0}$ varies
from $-1$ to $1$, the Chern number switches from $-1$ to $1$ [Fig.
\ref{fig:1}(e)], but the system never enters a trivial phase. {Correspondingly, the BIS evolves from an open hyperbolic contour in the continuum limit to a closed loop on the lattice. This closure arises because the compact, periodic nature of
the Brillouin zone effectively wraps the open hyperbolic branches
of the continuum model, forcing them to connect and form a closed
loop. } Specifically, the BISs
for $m_{0}=-1$ and $m_{0}=1$ form closed loops around the $X$
and $Y$ points, respectively [Fig. \ref{fig:1}(d,f)]. According to
recent theoretical advances in Chern number classification based on
high-symmetry points \cite{wan_classification_2025}, these two phases can be labeled as $C_{X}=-1$
and $C_{Y}=+1$, indicating that the topological charge is concentrated
at different points in the Brillouin zone. {We also discuss in the Appendix the form factors of other altermagnetic orders.
There we show that the sign reversal of the Dirac mass similarly drives a change in the Chern-number sign, accompanied by a transfer of the topological charge between different high-symmetry points in the Brillouin zone.}

In summary, Wilson and altermagnetic mass terms produce fundamentally
different effects when the Dirac mass changes sign. For the Wilson
mass, changing the Dirac mass sign drives a transition between trivial
and nontrivial Chern insulator phases, with the Chern number switching
from -1 to 0 via a band inversion at the $\Gamma$ point. In contrast,
for the altermagnetic mass, reversing the Dirac mass sign causes the
Chern number to switch directly between two nontrivial values of opposite
sign, relocating the topological charge between different high-symmetry
points without passing through a trivial phase. This unique transition
mechanism imparts distinctive properties to altermagnetic Weyl semimetals,
motivating the following introduction of a minimal altermagnetic Weyl
model.

\section{Minimal two bands model for Altermagnetic Weyl Semimetal}\label{III}

The close connection between 2D Chern insulators and 3D Weyl semimetals
provides a natural route to generalize the modified Dirac equation
to three dimensions. By promoting the Dirac mass in the 2D Dirac equation
to a $k_{z}$-dependent function, the 3D system can be viewed as a
series of 2D slices at fixed $k_{z}$, each described by a 2D Hamiltonian
with a $k_{z}$-dependent Dirac mass. In this framework, as $k_{z}$
is varied, the Chern number associated with each slice may change,
and Weyl nodes emerge at those critical $k_{z}$ where the gap closes,
signaling topological phase transitions between different 2D Chern
phases. Crucially, when the 2D model includes an altermagnetic mass
term, the interplay between the altermagnetic and Dirac masses produces
a novel topological structure unique to the altermagnetic Weyl semimetal.

Motivated by this picture, we construct the following minimal lattice
Hamiltonian for an altermagnetic Weyl semimetal
\begin{equation}
  \begin{aligned}
H_{min}(\mathbf{k})&=A\sin(k_{x})\sigma_{x}+A\sin(k_{y})\sigma_{y}\\
&+[m_{0}+2J(\cos k_{y}-\cos k_{x})+\cos k_{z}]\sigma_{z},
  \end{aligned}
\end{equation}
where $m_{0}$ is the Dirac mass parameter and $J$ controls the strength
of the altermagnetic mass. In our calculations, we take the parameters
$A=1$, $m_{0}=0$, and $J=1$. Here, the Dirac mass is explicitly
modified to depend on $k_{z}$ via the term $(m_{0}+\cos k_{z})\sigma_{z}$,
directly extending the 2D modified Dirac equation with altermagnetic
mass to three dimensions. This construction captures the essential
features of altermagnetic Weyl semimetals, including the unique interplay
between the $k_{z}$-dependent Dirac mass and the altermagnetic mass
term.

\begin{figure*}
\begin{centering}
\includegraphics[width=1.6\columnwidth]{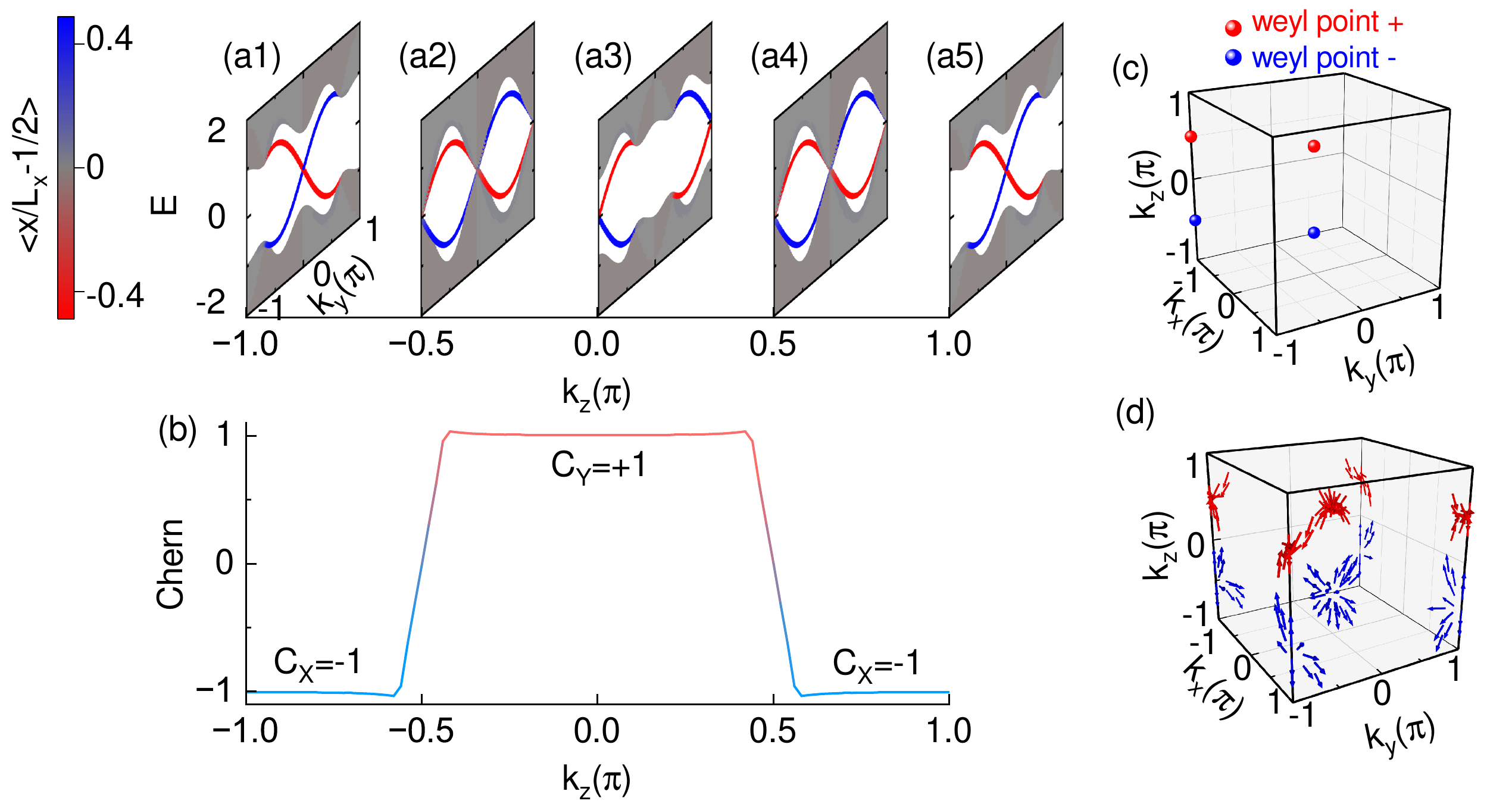}
\par\end{centering}
\caption{\label{fig:2}(a1\textendash a5) Slab energy spectra as a function
of $k_{z}$ in the Weyl semimetal phase, with eigenstates color-coded
according to their average position $\langle x\rangle$ to distinguish
surface and bulk states. (b) The corresponding Chern number $C(k_{z})$
for each fixed $k_{z}$ slice. (c) Distribution of Weyl nodes in the
Brillouin zone, with Fermi arcs connecting pairs of nodes. (d) Berry
curvature distribution in the Brillouin zone, highlighting regions
of concentrated topological charge.}
\end{figure*}

The hallmark of Weyl semimetals is the existence of topologically
protected surface states known as Fermi arcs, which connect the projections
of Weyl nodes of opposite chirality in the surface Brillouin zone \cite{wan_topological_2011,armitage_weyl_2018}.
To explicitly demonstrate these unique surface features and reveal
how they are modified by the altermagnetic mass, we discretize the
$x$ direction into $N=50$ lattice sites and compute the energy spectrum
$E(k_{y})$ as a function of $k_{z}$. Fig. \ref{fig:2}(a1\textendash a5) show representative spectra at $k_{z}=-\pi,-\pi/2,0,\pi/2,\pi$,
where each eigenstate is color-coded by its average position $\langle x\rangle$
to highlight its spatial character. States localized near the edges
correspond to surface states. Remarkably, for $|k_{z}|<\pi/2$, Fermi
arcs are found near $k_{y}=\pm\pi$, corresponding to chiral surface
states at the boundaries in the $x$ direction. For $|k_{z}|>\pi/2$,
the Fermi arcs shift to $k_{y}=0$, with their chirality reversed.
The chirality of these Fermi arc surface states is identified by the
direction of the group velocity surface states localized at opposite
edges propagate in opposite directions, as determined by the slope
of their energy dispersion. Unlike conventional Weyl semimetals, the
altermagnetic Weyl semimetal supports coexisting surface states with
opposite chirality on the same surface. This coexistence of counter-propagating
Fermi arcs is a hallmark of the altermagnetic Weyl phase, and originates
from the interplay between the altermagnetic mass and the Dirac mass.
Specifically, a sign reversal of the Dirac mass leads to a change
in the Chern number, which in turn produces Fermi arcs of opposite
chirality.

To further clarify the topological origin of the surface states, we
analyze the Chern number of the two-dimensional Hamiltonian $H_{min}(k_{x},k_{y},k_{z})$
as a function of $k_{z}$. For each fixed $k_{z}$, this Hamiltonian
reduces to a 2D modified Dirac model with an altermagnetic mass, as
discussed in Sec. \ref{II}. As illustrated in Fig. \ref{fig:2}(b), the surface Fermi
arcs arise from two distinct Chern insulating phases with different
Chern numbers in the 2D Brillouin zone. As $k_{z}$ varies, a topological
phase transition occurs at $|k_{z}|=\pi/2$:
for $|k_{z}|<\pi/2$, the BIS encloses the
X point, corresponding to a Chern number $C_{X}=-1$; for $|k_{z}|>\pi/2$,
the BIS encloses the Y point, and the Chern number switches to $C_{Y}=+1$ \cite{wan_classification_2025}.
This sharp change in Chern number is a direct manifestation of the
interplay between the Dirac mass and the altermagnetic mass. The transition
points $k_{z}=\pm\pi/2$ mark the locations of the Weyl nodes, situated
at $(k_{x},k_{y})=(0,0)$ and $(\pi,\pi)$ in the $k_{x}-k_{y}$ plane
[see Fig. \ref{fig:2}(c)]. Importantly, we find that the Fermi arc
connecting the two Weyl points at $(0,0,\pm\pi/2)$ exhibits chirality
opposite to that of the arc connecting $(-\pi,-\pi,\pm\pi/2)$. This opposite chirality of the fermi arcs is evident from both
the Chern number calculations and the surface state dispersions shown
in Fig. \ref{fig:2}(b).

To confirm the bulk\textendash boundary correspondence discussed above,
we computed the Berry-curvature distribution of the lower band over
the entire Brillouin zone, using a dense $200^{3}$ momentum grid
[Fig. \ref{fig:2}(d)]. For a generic two-band Hamiltonian $H(\mathbf{k})=\mathbf{d}(\mathbf{k})\!\cdot\!\boldsymbol{\sigma}$,
the Berry curvature reads \cite{shen_topological_2012}:
\begin{equation}
\boldsymbol{\Omega}(\mathbf{k})=\frac{1}{2}\,\frac{\mathbf{d}(\mathbf{k})}{|\mathbf{d}(\mathbf{k})|^{3}}\!\cdot\!\bigl(\partial_{k_{x}}\mathbf{d}\times\partial_{k_{y}}\mathbf{d}\bigr),
\end{equation}

which we evaluate numerically by finite differences. The curvature
is strongly peaked at the four Weyl nodes: positive monopoles at \((0,0,\pi/2)\) and \((-\pi,-\pi,\pi/2)\), and negative monopoles at \((0,0,-\pi/2)\) and \((-\pi,-\pi,-\pi/2)\). Integrating $\Omega_{z}(\mathbf{k})$
over each constant-$k_{z}$ slice reproduces the slice Chern numbers
obtained analytically, namely $C=+1$ for $|k_{z}|<\pi/2$ and $C=-1$
for $|k_{z}|>\pi/2$. This agreement provides a direct bulk confirmation
of the topological phase transition driven by the interplay between
the Dirac and altermagnetic mass terms.

\section{Multilayer Realization of Altermagnetic Weyl Semimetal}\label{IV}

\begin{figure}
\begin{centering}
\includegraphics[width=\columnwidth]{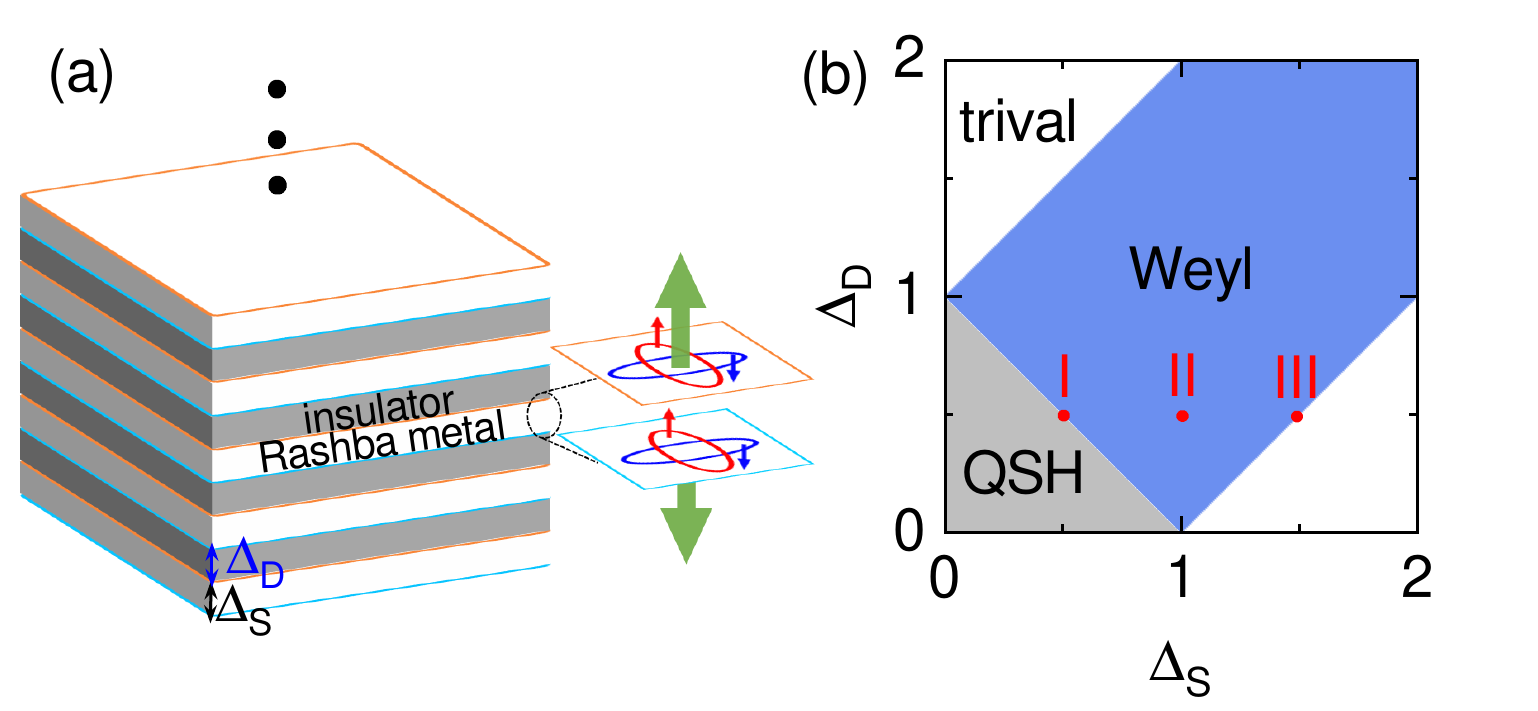}
\par\end{centering}
\caption{\label{fig:3}(a) Schematic illustration of the proposed altermagnetic
multilayer: two-dimensional metallic layers with $d$-wave altermagnetic
order are stacked along $z$, and adjacent layers carry Rashba spin\textendash orbit
couplings of opposite sign $(+\lambda,\,-\lambda)$, as shown by the large green arrows. Thin insulating
spacers suppress further-neighbour hopping, while intra-cell hybridization
$\Delta_{S}$ and inter-cell tunnelling $\Delta_{D}$ are indicated.
(b) Topological phase diagram as a function of the altermagnetic mass
$J$ and tunneling parameters $\Delta_{S}$, $\Delta_{D}$. The diagram
highlights the regions corresponding to the trivial insulator, Weyl
semimetal, and QSH phases. The three points (labeled I, II, III) indicate
parameter sets used for the subsequent bulk and edge calculations, with specific values $(\Delta_S, \Delta_D) = (0.5, 0.5)$ for I, (1, 0.5) for II, and (1.5, 0.5) for III. In all calculations, we set $t=0.1$, $\lambda=0.5$, and $J=0.5$.}
\end{figure}

To illustrate that the same mass\textendash interplay mechanism can
arise in a completely different setting, we consider a synthetic multilayer
of two-dimensional $d$-wave altermagnetic metals.
As sketched in Fig. \ref{fig:3}(a), neighbouring layers are engineered to have opposite Rashba spin-orbit coupling strengths $+\lambda$ and $-\lambda$, while thin
insulating spacers suppress direct hopping beyond nearest neighbours
along $z$ \cite{burkov_weyl_2011}.

In momentum space the stack Hamiltonian is described by (lattice constant
is set to unity)
\begin{widetext}
\begin{equation}
\begin{aligned}
H({\bf k}_{\parallel}) & =\sum_{i,j}c_{{\bf k}_{\parallel},i}^{\dagger}\!\Bigl[-t\,(\cos k_{x}+\cos k_{y})\sigma_{0}\tau_{0}\delta_{ij}
+\lambda\,\tau_{z}\bigl(\sin k_{x}\,\sigma_{x}+\sin k_{y}\,\sigma_{y}\bigr)\delta_{ij}+J\bigl(\cos k_{y}-\cos k_{x}\bigr)\sigma_{z}\tau_{0}\delta_{ij}\\
 & +\Delta_{S}\,\sigma_{0}\tau_{x}\delta_{ij}
 +\tfrac{\Delta_{D}}{2}\sigma_{0}\bigl(\tau_{+}\delta_{j,i+1}+\tau_{-}\delta_{j,i-1}\bigr)\Bigr]c_{{\bf k}_{\parallel},j},
\end{aligned}
\end{equation}
\end{widetext}
Here $c_{{\bf k}_{\parallel},i}^{\dagger}$ creates an electron with in-plane
momentum ${\bf k}_{\parallel}=(k_{x},k_{y})$ in layer $i$; $\sigma_{\alpha}$
act on real spin, whereas $\tau_{\alpha}$ act on the layer degree
of freedom and  $\tau_{\pm} = (\tau_x \pm i\tau_y)/2$. The first term is the isotropic kinetic energy. The $\lambda$ term encodes the layer-alternating Rashba spin–orbit coupling, which can naturally arise in a multilayer structure due to opposite electrostatic potential gradients at the top and bottom surfaces of each metallic layer, resulting in opposite Rashba spin-orbit coupling for adjacent layers.{ As a practical note, a similar Rashba-type spin texture can also be realized in thin films of three-dimensional topological insulators, whose top and bottom surfaces possess opposite spin orientations; when coupled to altermagnetic layers, such TI films can effectively provide the alternating Rashba coupling $(+\lambda, -\lambda)$ required in our multilayer structure.
The $J$ term is the $d$-wave altermagnetic exchange, which can be realized in materials such as $\mathrm{RuO}_2$ \cite{ARPESRuO2} and $\mathrm{KV}_2\mathrm{Se}_2\mathrm{O}$ \cite{ARPESKV2Se2O} where momentum-dependent spin splitting has been observed in recent ARPES measurements.} The last two terms describe, respectively, intra-cell hybridisation ($\Delta_{S}$) and inter-cell hopping ($\Delta_{D}$) along $z$. After Fourier transforming in the stacking direction,
$\Delta_{S}$ and $\Delta_{D}$ combine into a $k_{z}$-dependent
mass, $m(k_{z})=\Delta_{S}-\Delta_{D}\cos k_{z}.$

It is convenient to block-diagonalize the Hamiltonian by a unitary
transformation
\begin{equation}
U=\exp\left(i\frac{\pi}{4}\sigma_{z}\tau_{z}\right).
\end{equation}
This transformation mixes the spin and pseudospin sectors such that
\begin{equation}
U\,\tau_{\pm}U^{\dagger}=\sigma_{z}\tau_{\pm}.
\end{equation}

Physically, this rotates the basis so that the coupling between layers, which encoded in $\tau_{\pm}$ terms, becomes an effective $\sigma^{z}$
term in each block. After applying $U$, the Hamiltonian decouples
into two $2\times2$ blocks . The transformed Hamiltonian can be
written as
\begin{equation}
\begin{aligned}
\mathcal{H}'(\mathbf{k}) & =-t\,(\cos k_{x}+\cos k_{y})\sigma_{0}+\lambda(\sin k_{x}\,\sigma_{x}+\sin k_{y}\,\sigma_{y})\;\\
 & +\;\Big[\,J(\cos k_{y}-\cos k_{x})+\hat{\Delta}(k_{z})\,\Big]\sigma_{z}\,,
\end{aligned}
\end{equation}
where $\hat{\Delta}(k_{z})$ can be replaced by its eigenvalues $\pm\Delta(k_{z})$,
with
\begin{equation}
\Delta(k_{z})=\sqrt{\Delta_{S}^{2}+\Delta_{D}^{2}+2\,\Delta_{S}\Delta_{D}\cos(k_{z}d)}\,.\label{Delta}
\end{equation}
where $d$ is the superlattice period along the growth ($z$) direction. We have thereby reduced the problem to a family of
$2\times2$ Bloch Hamiltonians parameterized by $k_{z}$, each of
the form
\begin{equation}
  \begin{aligned}
H_{2D}(k_{x},k_{y};k_{z})&=-t(\cos k_{x}+\cos k_{y})\sigma_{0}+\lambda(\sin k_{x}\sigma_{x}\\
&+\sin k_{y}\sigma_{y})+M(\mathbf{k})\sigma_{z},
  \end{aligned}
\end{equation}
with an effective mass term
\begin{equation}
M(\mathbf{k})=J(\cos k_{y}-\cos k_{x}) \pm \Delta(k_{z})\,.
\end{equation}

This $2\times2$ Hamiltonian is formally equivalent to a modified
Dirac equation with altermagnetic mass, where $\Delta(k_{z})$ serves
as a $k_z$-dependent Dirac mass term. The bulk spectrum is given by
\begin{equation}
E(\mathbf{k})=-t\,(\cos k_{x}+\cos k_{y})\pm\sqrt{\lambda^{2}(\sin^{2}k_{x}+\sin^{2}k_{y})+M(\mathbf{k})^{2}}.
\end{equation}

Bulk gap closings require $\sin k_{x}=\sin k_{y}=0$ and $M(\mathbf{k})=0$
are satisfied. The first condition restricts $(k_{x},k_{y})$ to $(0,0),\ (0,\pi),\ (\pi,0),\ (\pi,\pi)$.
At $(0,0)$ and $(\pi,\pi)$, since $\cos k_{y}-\cos k_{x}=0$, the
mass term reduces to $M=\pm \Delta(k_{z})$. Thus, gap closing requires $\Delta(k_{z})=0$. According to Eq. (\ref{Delta}), the condition is given by
\begin{equation}
\cos(k_{z}d)=-\frac{\Delta_{S}^{2}+\Delta_{D}^{2}}{2\Delta_{S}\Delta_{D}}.
\end{equation}
A real solution for $k_{z}$ exists only when the right-hand side
is within $[-1,1]$, i.e., when $\Delta_{S}=\Delta_{D}$. Therefore,
\textbf{f}or generic parameters, these degeneracies are lifted and
the system remains gapped at these points.

At the points $(0,\pi)$ and $(\pi,0)$, where $\cos k_{y}-\cos k_{x}=\pm2$,
the mass term becomes $M=\pm2J \pm \Delta(k_{z})$. The gapless condition
then reads
\begin{equation}
\Delta(k_{z})=\mp2J,\label{eq:weyl}
\end{equation}
which gives
\begin{equation}
\Delta_{S}^{2}+\Delta_{D}^{2}+2\Delta_{S}\Delta_{D}\cos(k_{z}d)=4J^{2},
\end{equation}
and hence
\begin{equation}
\cos(k_{z}d)=\frac{4J^{2}-(\Delta_{S}^{2}+\Delta_{D}^{2})}{2\Delta_{S}\Delta_{D}}.
\end{equation}

The existence of real solutions requires
\begin{equation}
-1\leq\frac{4J^{2}-(\Delta_{S}^{2}+\Delta_{D}^{2})}{2\Delta_{S}\Delta_{D}}\leq1,
\end{equation}
or equivalently,
\begin{equation}
\frac{(\Delta_{S}-\Delta_{D})^{2}}{4}\leq J^{2}\leq\frac{(\Delta_{S}+\Delta_{D})^{2}}{4}.
\end{equation}
Introducing the notations
\begin{equation}
J_{c1}=\frac{|\Delta_{S}-\Delta_{D}|}{2},\qquad J_{c2}=\frac{\Delta_{S}+\Delta_{D}}{2},
\end{equation}
we find that Weyl nodes exist only within the parameter window
\begin{equation}
J_{c1}\leq J\leq J_{c2}.
\end{equation}

In this regime, Eq. (\ref{eq:weyl}) admits two real solutions for each
of the points $(0,\pi)$ and $(\pi,0)$,
corresponding to distinct values of $k_{z}$. Thus, a total of four
Weyl nodes emerge at positions $(0,\pi,\pm k_{z}^{c})$ and $(\pi,0,\pm k_{z}^{c})$,
each set forming a pair of nodes with opposite chirality, as required
by the Nielsen\textendash Ninomiya theorem \cite{,nielsen_no-go_1981,nielsen_adler-bell-jackiw_1983}. Here, $k_z^c$ is given by
\[
k_z^c = \frac{1}{d} \arccos \left( \frac{4J^2 - (\Delta_S^2 + \Delta_D^2)}{2\Delta_S\Delta_D} \right).
\]

With the above analysis, we clearly outline the topological phase
diagram at fixed altermagnetic exchange parameter $J=0.5$, as shown
in Fig.\ref{fig:3}(b). The diagram features three distinct topological
regions, defined by the tunneling parameters $\Delta_{S}$ and $\Delta_{D}$.

For sufficiently large differences ($|\Delta_{S}-\Delta_{D}|>2J$),
the system is fully gapped and trivial, as dominant tunneling prevents
band inversion. Conversely, in the intermediate regime ($|\Delta_{S}-\Delta_{D}|<2J\leq\Delta_{S}+\Delta_{D}$),
Weyl nodes appear, forming a Weyl semimetal phase. These nodes persist
until the critical boundaries ($|\Delta_{S}-\Delta_{D}|=2J$ or $\Delta_{S}+\Delta_{D}=2J$)
where they annihilate pairwise at the Brillouin zone edge, signaling
a topological transition.

Finally, when $\Delta_{S}+\Delta_{D}<2J$, the system transitions
to a 3D quantum spin Hall (QSH) insulator phase, with robust helical
surface states throughout the Brillouin zone. This phase is distinct
from the trivial insulator and Weyl semimetal phases.

\begin{figure*}
\begin{centering}
\includegraphics[width=1.6\columnwidth]{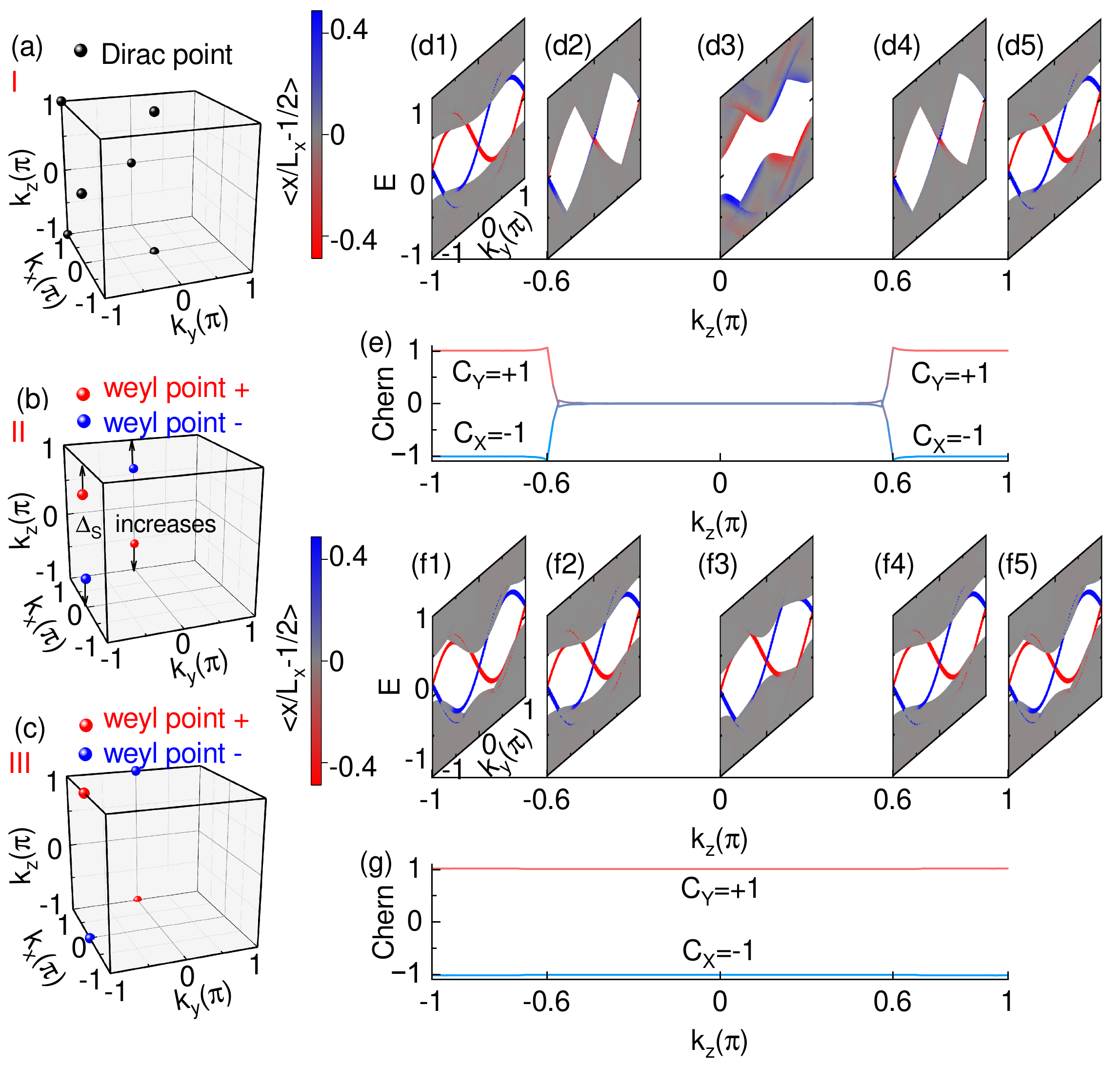}
\par\end{centering}
\caption{\label{fig:4}(a\textendash c) Fermi surface distributions for three representative parameter sets, marked by points I, II, and III in Fig. \ref{fig:3}(b). (d1\textendash d5) Slab spectra
as a function of $k_{z}$ in the Weyl semimetal phase with $\Delta D=0.5$ and $\Delta S=1$, with eigenstates
color-coded by their average position $\langle x\rangle$. (e) The
corresponding Chern number $C(k_{z})$, which exhibits a jump at the
Weyl nodes. (f1\textendash f5) Slab spectra as a function of $k_{z}$
in the QSH phase with $\Delta D=0.5$ and $\Delta S=0.2$, again color-coded by average position. (g) The corresponding
Chern number $C(k_{z})$, remaining constant throughout, consistent
with robust QSH edge states and the absence of Weyl nodes.}
\end{figure*}


Figure \ref{fig:4} demonstrates the topological evolution across these phases as the intra-layer tunneling parameter $\Delta_{S}$ is varied. Panels (a)–(c) show the Fermi surface contours for the three representative parameter sets [corresponding to points I–III in Fig. \ref{fig:3}(b)], directly visualizing the sequence from the 3D QSH phase, through the Weyl semimetal, to the trivial insulator.

In the Weyl semimetal phase {[}Fig. \ref{fig:4}(b){]}, the bulk band
structure hosts two pairs of Weyl nodes located at $(k_{x},k_{y},k_{z})=(0,-\pi,\pm k_{z}^{c})$
and $(\pi,0,\pm k_{z}^{c})$, where $k_{z}^{c}$ denotes their positions
along the $k_{z}$-axis. The separation and location of these Weyl
nodes are primarily controlled by $\Delta_{S}$. As $\Delta_{S}$
increases, the Weyl nodes move toward the Brillouin zone boundaries
at $k_{z}=\pm\pi${[}Fig.
\ref{fig:4}(c){]}, eventually annihilating in pairs of opposite chirality
at these boundaries, driving the system into a trivial insulating
phase. Conversely, decreasing $\Delta_{S}$ shifts the Weyl nodes
toward the $k_{z}=0$ plane. At a critical $\Delta_{S}$ value {[}Fig.
\ref{fig:4}(a){]}, the Weyl nodes merge into Dirac points at $(0,-\pi,0)$
and $(\pi,0,0)$, and additional gapless points emerge simultaneously
at other high-symmetry momenta $(0,0,\pm \pi)$ and $(\pi,-\pi,\pm \pi)$.
Further decreasing $\Delta_{S}$ reopens a full bulk gap without any
Weyl nodes, leading the system into a three-dimensional quantum spin
Hall insulating phase characterized by robust topological surface
states, as discussed below.

Here, a key signature of the altermagnetic Weyl semimetal is the emergence
of helical Fermi arcs on the side surfaces, in contrast to the single
chiral arc seen in conventional Weyl semimetals \cite{wan_topological_2011,armitage_weyl_2018}. {In this context, the term helical Fermi arcs denotes a pair of counter-propagating surface states with opposite group velocities on the same surface, rather than spin–momentum locking.} To demonstrate this,
we perform a slab calculation, taking the system finite along the
$x$ direction and infinite along $y$ and $z$. The calculated surface
band structure {[}Fig. \ref{fig:4}(d1)\textendash (d5){]} shows that,
between the two Weyl nodes at $k_{z}=\pm k_{z}^{c}$, the bulk gap
remains open, but in-gap states localized on the surfaces appear.
These surface states, color-coded by position, form Fermi arcs that
traverse the gap and connect the projections of the Weyl nodes. Crucially, on the same surface, Fermi arcs with opposite group velocities coexist, so their net chirality vanishes. The emergence of such helical Fermi arcs arises from the interplay between the Dirac mass and the altermagnetic mass. Specifically, the two blocks of the Hamiltonian share the same altermagnetic mass but have Dirac masses of opposite sign, leading to opposite Chern numbers in the two blocks. As a result, on the surface, coexisting Fermi arcs with opposite chirality are generated, which is a hallmark feature of the altermagnetic Weyl phase.

This momentum-space distribution of Fermi arcs is intimately linked
to the partial Chern numbers defined at high-symmetry points in each
$k_{z}$ slice. For every fixed $k_{z}$, the effective two-dimensional
Hamiltonian takes the form of a modified Dirac equation with an altermagnetic
mass term. Due to inversion symmetry, the topological structure can
be further refined by considering skyrmion charges at the high-symmetry
points $(0,\pi)$ and $(\pi,0)$ \cite{wan_classification_2025}.
Considering that the Dirac mass terms have opposite sign in the two blocks of the Hamiltonian, their
interplay with the altermagnetic mass leads to partial Chern numbers
$C_{X}=-1$ and $C_{Y}=+1$ at the respective high-symmetry points.
This structure directly dictates the existence of counter-propagating
Fermi arcs: boundary modes associated with $C_{X}=-1$ appear near
$k_{y}\approx0$, while those with $C_{Y}=+1$ localize near $k_{y}\approx\pi$.
The opposite signs set their chirality, and the association with different
high-symmetry points ensures the robustness and coexistence of these
paired arcs on the same surface against long-range disorder\cite{wan_classification_2025,LI20241221}. { In particular, after the
unitary transformation $U=e^{i\pi\sigma_z\tau_z/4}$, the Hamiltonian
becomes block-diagonal, revealing an emergent layer–spin locking symmetry
that ties the spin orientation to the layer index and prevents hybridization
between the two sectors in the ideal limit. Moreover, even when a weak
inter-block coupling is introduced, the Fermi arcs remain gapless because
they originate from distinct high-symmetry points carrying localized
topological charges ($C_X=-1$, $C_Y=+1$) that cannot annihilate or
mix. Therefore,both of the layer–spin locking symmetry and
high-symmetry-point Chern protection ensure the stability of the helical
surface states.
}
The evolution of these topological features as a function of $k_{z}$
is further reflected in the calculated Chern number $C(k_{z})$ for
each two-dimensional $k_{x}$\textendash $k_{y}$ slice {[}Fig. \ref{fig:4}(e){]}.
In the Weyl phase, $C(k_{z})$ undergoes quantized jumps of $\pm1$
as $k_{z}$ crosses each Weyl node, but the total Chern number always
vanishes because the two blocks of the Hamiltonian experience opposite
changes. This distribution of partial Chern numbers underlies the
coexistence of helical Fermi arcs and defines the unique topological
nature of the altermagnetic Weyl semimetal.

Upon further decreasing $\Delta_{S}$, the Weyl nodes move toward the $k_z = 0$ plane and eventually annihilate, driving the system into a three-dimensional QSH) phase [Fig. \ref{fig:4}(a)]. In this regime, the Chern numbers of each block remain fixed for all $k_z$ [Fig. \ref{fig:4}(g)], and robust helical edge states persist throughout the Brillouin zone [Figs. \ref{fig:4}(f1)–\ref{fig:4}(f5)]. Thus, by
tuning system parameters, a sequence of topological phase transitions
is induced, with the interplay between the Dirac and altermagnetic
masses giving rise to the emergence of paired Weyl nodes, helical
Fermi arcs, and ultimately a strong QSH phase. This progression highlights
the unique bulk-boundary correspondence and the rich topological phase
diagram enabled by altermagnetic order.

\section{Summary}\label{V}

In summary, we have demonstrated that the introduction of an altermagnetic
mass in a modified Dirac Hamiltonian leads to a distinctive Weyl semimetal
phase, fundamentally different from the conventional Wilson mass scenario.
The altermagnetic mass drives transitions between nontrivial topological
phases by shifting the location of topological charge among high-symmetry
points, resulting in pairs of Weyl nodes and the coexistence of helical
Fermi arcs with opposite chirality on the same surface. Our minimal
lattice model, supported by explicit calculations of bulk and surface
spectra, Berry curvature, and Chern numbers, reveals a rich topological
phase diagram including trivial insulator, Weyl semimetal, and quantum
spin Hall phases. Finally, we propose that multilayer structures composed
of 2D Rashba metal and insulating spacers offer a
platform for realizing altermagnetic Weyl semimetals. Our results
illuminate the rich topological physics accessible via altermagnetism,
clarify the unique impact of the altermagnetic mass term on Dirac
band topology, and provide concrete guidelines for the experimental
realization and detection of altermagnetic Weyl semimetals.

\section*{acknowledgements}

This work was financially supported by the National Key R and D Program
of China (Grant No. 2024YFA1409002), the National Natural Science
Foundation of China (Grants No. 12374034, Grants No. 124B2069, and Grants No. 12447147), the Quantum Science and Technology-National Science and Technology Major Project (Grant No. 2021ZD0302403), and the China Postdoctoral Science Foundation (Grant No. 2024M760070). The computational resources are supported by the High-Performance
Computing Platform of Peking University.

{

\section*{Appendix: \(\mathcal{C}_6\mathcal{T}\)-symmetric altermagnetic mass and Dirac topology}
Recent theoretical and experimental studies have identified MnTe as a representative \(g\)-wave altermagnet \cite{krempasky_altermagnetic_2024}, whose momentum-dependent Zeeman splitting in the \(k_{z}>0\) sector respects the combined \(\mathcal{C}_{6}\mathcal{T}\) symmetry. By placing a two-dimensional Rashba metal in proximity to such a \(g\)-wave altermagnet, one can induce an effective \(\mathcal{C}_{6}\mathcal{T}\)-symmetric altermagnetic mass term via interfacial exchange coupling or magnetic proximity effect. Notably, this effective mass term takes the same functional form as the two-dimensional \(f\)-wave magnetic order, which can naturally exist on hexagonal-like lattices \cite{fwave}.

To directly examine the topological effect of the \(\mathcal{C}_{6}\mathcal{T}\)-symmetric altermagnetic mass, we consider a Dirac Hamiltonian on a hexagonal lattice with a conventional Dirac mass \(m_{0}\) and an altermagnetic mass of strength \(J\):

\begin{equation}
\begin{aligned}
H^{h}(\mathbf{k}) =
\sum_{i=1}^{3} v_{F} \sin\left(\mathbf{k}\!\cdot\!\mathbf{a}_{i}\right)
\big[(\hat{\mathbf{a}}_{i}\!\cdot\!\hat{x})\sigma_{x}
+(\hat{\mathbf{a}}_{i}\!\cdot\!\hat{y})\sigma_{y}\big]\\
+\Big\{ m_{0}+J
\!\left[\sum_{i=1}^{3}\sin\left(\mathbf{k}\!\cdot\!\mathbf{a}_{i}\right)\right]\!\Big\}\sigma_{z}.
\end{aligned}
\end{equation}

Here, the nearest-neighbor bond vectors are
\(\mathbf{a}_{1}=(1,0)\),
\(\mathbf{a}_{2}=(-1,\sqrt{3})/2\),
and \(\mathbf{a}_{3}=(-1,-\sqrt{3})/2\)
(the lattice constant is set to unity). 

For this two-band Hamiltonian \(H(\mathbf{k})=\mathbf{d}(\mathbf{k})\!\cdot\!\boldsymbol{\sigma}\),
the vector \(\mathbf{d}(\mathbf{k})=(d_{x},d_{y},d_{z})\)
defines a mapping from the Brillouin zone to the Bloch sphere,
which encodes the topological character of the band structure and is directly related to the Chern number.

As shown in Fig.\ref{fig:R1}(a,b),
the \(\mathbf{d}(\mathbf{k})\) textures for opposite mass signs (\(m_{0}=\pm1\))
exhibit band-inversion surfaces enclosing the \(K\) and \(K^{\prime}\) points, respectively.
From these textures, one finds that the two Dirac sectors carry opposite Chern numbers,
\(C_{K}=+1\) and \(C_{K^{\prime}}=-1\),
confirming that the sign reversal of the Dirac mass drives a Chern-number inversion—
exactly as in the \(d\)-wave case discussed in the main text.
This demonstrates that the same topological mechanism applies to altermagnetic masses with different form factors,
including those with \(\mathcal{C}_{6}\mathcal{T}\) symmetry.

\begin{figure}[t]
\includegraphics[width=0.9\columnwidth]{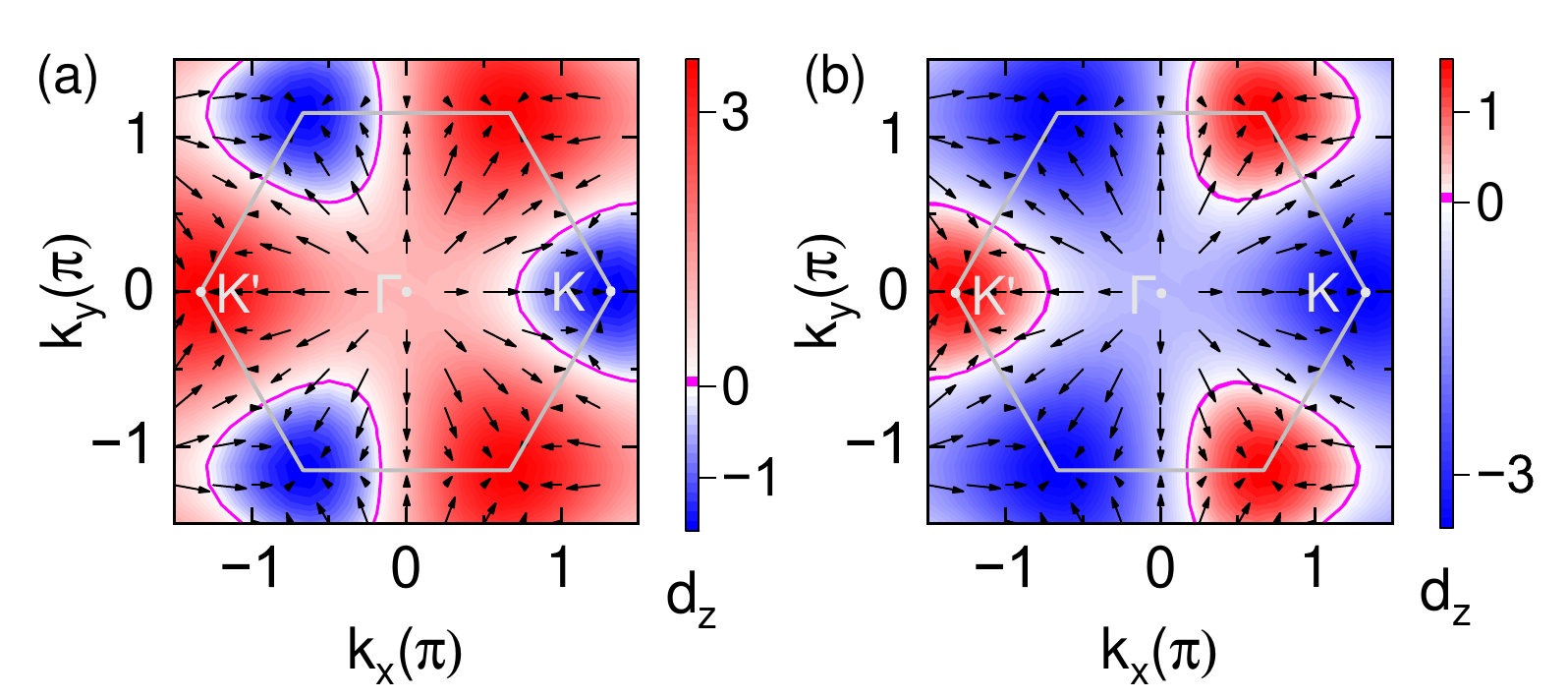}
\caption{\textbf{(a,b)}  Momentum-space textures of
\(\mathbf{d}(\mathbf{k})=(d_x,d_y,d_z)\)
for the hexagonal-lattice model with Dirac masses \(m_0=\pm1\),
\(v_F=1\), and \(J=1\).
The band-inversion surfaces defined by \(d_z(\mathbf{k})=0\)
enclose the \(K\) and \(K^{\prime}\) valleys for opposite mass signs}
\label{fig:R1}
\end{figure}
}

\bibliography{alterweyl.bib}

\end{document}